\documentclass[nofootinbib,onecolumn]{revtex4}
\usepackage{graphicx}
\usepackage{subfig}
\usepackage{amssymb}
\usepackage{amsmath}
\usepackage{bm}
\usepackage{slashbox}

\begin{document}
\title{Braneworld gravity within non-conservative gravitational theory}
%\date{\today}
\author{J. C. Fabris}
\email{julio.fabris@cosmo-ufes.org}
\author{Thiago R. P. Caram\^es}
\email{trpcarames@gmail.com}
\affiliation{N\'ucleo Cosmo-ufes \& Departamento de F\'isica, CCE, Universidade Federal do Esp\'irito Santo (UFES)\\
Vit\'oria-ES-Brazil}
\author{J. M. Hoff da Silva}
\email{hoff@feg.unesp.br}
\affiliation{Departamento de F\'isica e Qu\'imica, Universidade Estadual Paulista (UNESP),\\
 Guaratinguet\'a-SP-Brazil}


\begin{abstract} 
We investigate the braneworld gravity starting from the non-conservative gravitational field equations in a five-dimensional bulk. The approach is based on the Gauss-Codazzi formalism along with the study of the braneworld consistency conditions. The effective gravitational equations on the brane are obtained and the constraint leading to a brane energy-momentum conservation is analyzed. 
\end{abstract}

\maketitle
\noindent

\section{Introduction}

Despite the ubiquity of dissipative processes in the real world, it is intriguing to notice their absence in the standard formulations of the principle of least action. In the traditional classical mechanics, dissipative phenomena are handled by means of the Rayleigh dissipation function which comes into play through an extra term in the Euler-Lagrange equations, where one does not abandon however the underlying variational formalism, so that the Lagrangian of the system is kept untouched \cite{goldstein}. A first effort attempting to construct, within the classical mechanics context, a full formalism to describe dissipative systems from the perspective of a principle of least action dates back to the Herglotz's work \cite{herglotz}. In his approach he argue that it would be possible to describe a physical system endowed with dissipation by assuming an action-dependent Lagrangian. For instance, when considering a linear dependence on the action he has shown the appearance of a typical velocity-proportional frictional term in the corresponding equations of motion derived from such Lagrangian. Almost ninety years separated the pioneer Herglotz's contribution from a covariant extension of his formalism, which was just recently accomplished by Lazo {\it et al} \cite{lazo}. From this covariant formulation the authors constructed a new theory of gravity consisting of a set of modified field equations along with a non-conservation for the energy-momentum tensor. They make a discussion on the possible consequences of this ``geometric'' dissipation effects on the cosmological scenario, pointing out that these new degrees of freedom can account for the dark energy content in the universe. Besides, they add an study on the gravitational waves propagation within this theory. 

In a very recent paper, a more complete investigation of cosmological aspects in this non-conservative gravity is performed \cite{ijmpd}. At the background level, the authors show an equivalence between this non-conservative cosmology with the bulk viscous model in the Eckart's formalism \cite{winfried,roy}. Whereas at the perturbative level, they verified that the linear perturbations indicate a possible way out to alleviate the problems faced by the viscous cosmology.       

Braneworld models have attracted the attention of the scientific community due its possible application to the hierarchy problem \cite{RSI}. Soon after the appearance of such a possibility, the gravitational aspects of these models started to be under investigation. In particular, a systematic study performed by means of the Gauss-Codazzi formalism \cite{SSM,ALI} made possible a broad range of applications of braneworld scenarios in gravitation and cosmology. From among the several interesting prospects resulting from this investigation, in the context of braneworld gravity, is the impossibility of covariant conservation of the brane stress tensor when matter is present in the bulk \cite{Maa}. The main purpose of this paper is to investigate the physical consequences of such a geometric induced non-conservation of the energy-momentum on the construction of braneworld models.

After a short introduction highlighting some of the main aspects of non-conservative gravity in Section II, we apply in Section III the Gauss-Codazzi formalism assuming that the bulk gravity is governed by its precepts. It is shown that the non-conservative aspect of the bulk gravity can counterbalance the bulk matter effect leading to a covariant conservation of the brane stress-tensor. It is also shown a complete gravitational effective field equation, along with a corrected four-dimensional gravitational 'constant', which now acquires a dependence upon the coordinates. In Section IV we approach non-conservative braneworld models with the aid of the well known braneworld sum rules, a complete formalism resulting in a one-parameter family of consistency conditions. It is shown that, in this specific context, it is possible to derive an extension of the Randall-Sundrum model without using a negative brane tension. In the final Section we conclude emphasizing the possible applications in cosmology.

% by first principles  is one of the most remarkable conceptual challenges faced by the theoretical physics.

\section{A toolkit on non-conservative gravity}

As discussed in the introduction in the Ref. \cite{herglotz} G. Herglotz realized the possibility of incorporating dissipative systems into the principle of least action by means of a action-dependent Lagrangian as follows
\begin{eqnarray}
\label{herglotz}
S = \int {\cal L}(x, \dot x, S) d t,
\end{eqnarray}
where $x=x(t)$ denotes the path that extremises the action $S$. Such a condition leads to a generalized version of the Euler-Lagrangian equation
\begin{eqnarray}
\label{HEL}
\frac{d}{dt}\frac{\partial {\cal L}}{\partial \dot x} - \frac{\partial {\cal L}}{\partial x}  - \frac{\partial{\cal L}}{\partial S}\frac{\partial {\cal L}}{\partial \dot x} = 0.
\end{eqnarray}
We can illustrate how the description of dissipative processes can emerge from (\ref{HEL}) by considering a simple case where the Lagrangian has a linear dependence in the action ${\cal L}=\frac{m\dot{x}^2}{2}-U(x)-\frac{\gamma}{m} S$, which results in a equation of motion endowed with a typical friction term $\gamma \dot{x}$. In the covariant generalization of this formalism introduced in \cite{lazo}, the authors proposes a gravitational theory given by the following extended Einstein-Hilbert lagrangian
\begin{eqnarray}
\label{ADL}
{\cal L} = \sqrt{-g}(R - \lambda_{\mu}s^\mu) + {\cal L}_m,
\end{eqnarray}
where $s^\mu$ denotes an action-density field, while $\lambda_\mu$ is a parameter encoding the emerging dissipative effects. According to what is shown in \cite{lazo}, $s^{\mu}$ shall disappear during the extremisation of the action, so that it does not show up in the field equations. The coupling four-vector $\lambda^{\mu}$ may be in general coordinate-dependent, although we shall concentrate in the simplest case, in which its components are constant. The term $\lambda_{\mu}s^\mu$ can be interpreted as a covariant generalization of the classical linear action-dependence mentioned a few lines above\footnote{The Ref.\cite{ijmpd} brings a more detailed discussion about the choice (\ref{ADL}).}. 
%This can be easily seen throug the detailed steps performed in \cite{lazo}. In such Ref., the authors proceed to the covariant extension of Herglotz approach by looking for a covariant version of the time derivative of the action appearing in its Eq. (1). They accomplish it by assuming this time derivative as non-covariant version of a four-divergence of a certain four vector $s^{\mu}$ for which the term indexed as $\mu=0$ means just the classical time derivative of the action one wanted to generalize. 
This non-conservative theory of gravity presents the following set of field equations
\begin{equation}
^{(5)}G_{\mu\nu}(R)+\,^{(5)}G_{\mu\nu}({\cal K})=\kappa_{5}^{2}T_{\mu\nu},\label{1}
\end{equation} where $\kappa_{5}^{2}$ is the gravitational coupling constant in five dimensions, $\,^{(5)}G_{\mu\nu}(\xi)$ is the Einstein tensor associated to the tensorial quantity $\xi$, and ${\cal K}_{\mu\nu}=\lambda_{\alpha}\Gamma^{\alpha}_{\mu\nu}-\frac{1}{2}\Big(\lambda_{\nu}\Gamma^{\alpha}_{\mu\alpha}+\lambda_{\mu}\Gamma^{\alpha}_{\nu\alpha}\Big)$. The five-vector $\lambda_\mu$ is the responsible for the geometric non-conservation, since the covariant divergence of (\ref{1}) shall not vanish for a non-null $\lambda_\mu$. 

\section{Applying the Gauss-Codazzi formalism}

As already remarked in the Introduction, we shall start assuming a five-dimensional bulk whose gravitational interaction is governed by non-conservative gravity, i. e., in five dimensions the field equation is given by Eq. (\ref{1}). In order to project the geometric relevant quantities on the brane and find the effective gravitational equation in four dimensions, we implement the well known Gauss-Codazzi formalism, starting from Gauss equation
\begin{equation}
^{(4)}R^{\alpha}_{\beta\gamma\delta}=\,^{(5)}R^\mu_{\nu\rho\sigma}q_\mu^\alpha q_\beta^\nu q_\gamma^\rho q_\delta^\sigma + K^\alpha_\gamma K_{\beta\delta}- K^\alpha_\delta K_{\beta\gamma},\label{gauss}
\end{equation} relating (projecting) the five-dimensional curvature to its four-dimensional counterpart. The tensor $K_{\mu\nu}$ is the extrinsic curvature. Regarding Eq. (\ref{gauss}) some words are in order. The five-dimensional line element is understood as 
\begin{equation}
ds^2=q_{\mu\nu}dx^\mu dx^\nu+dr^2, \label{le}
\end{equation} 
where $q_{rr}=0$ and $r$ is the index of the fifth dimension. Besides, one denotes $g_{\mu\nu}=q_{\mu\nu}+n_\mu n_\nu$, where $n_\mu$ is a unitary vector orthogonal everywhere on the brane, provided it is orientable. In terms of (the variation of) $n_\mu$ the extrinsic curvature reads $K_{\mu\nu}=q_\mu^\alpha q_\nu^\beta \nabla_\alpha n_\beta$. It is clear from these choices that from $0$ to $4$ in the indexes we are restricted to the brane, leaving the last index value to the extra dimension. Notice that the physical content of Eq. (\ref{gauss}) may be simply stayed as follows: the brane curvature is given by the projection of the bulk curvature, also having into account the way the brane is embedded in the bulk. 

After some manipulation it is possible to write (\ref{gauss}) as 
\begin{equation}
^{(4)}G_{\beta\delta}(R)=\,^{(5)}R_{\nu\sigma}q_\beta^\nu q_{\delta}^\sigma -\frac{1}{2}q_{\beta\delta}\,^{(5)}R_{\nu\sigma}q^{\nu\sigma}+K K_{\beta\delta}- K_{\alpha\beta} K^\alpha_\delta-\frac{1}{2}(K^2-K_{\alpha\gamma}K^{\alpha\gamma})q_{\beta\delta}+\frac{1}{2}q_{\beta\delta}\,^{(5)}R^\mu_{\nu\rho\sigma}n_\mu n^\rho q^{\sigma\nu}-\tilde{E}_{\beta\delta},\label{pro1}
\end{equation} where $\tilde{E}_{\beta\delta}=\,^{(5)}R^\mu_{\nu\rho\sigma}n_\mu n^\rho q_\beta^\nu q_\delta^\sigma$. Directing the calculation to make contact with the approaches existing in the literature, we shall make use of the five-dimensional Weyl tensor, $C^\alpha_{\beta\rho\delta}$, along with usual algebraic manipulations in terms of what we have 
\begin{eqnarray}
^{(4)}G_{\beta\delta}(R)&=&\left.\,^{(5)}G_{\nu\sigma}(R)q_\beta^\nu q_\delta^\sigma+K K_{\beta\delta}- K_{\alpha\beta}K^\alpha_{\delta}-\frac{1}{2}(K^2-K_{\alpha\gamma}K^{\alpha\gamma})q_{\beta\delta}-E_{\beta\delta}-\frac{1}{3}\,^{(5)}R_{\sigma\nu}q^\nu_\beta q^\sigma_\delta\right.\nonumber\\&-&\left.\frac{2}{3}q_{\beta\delta}\Big[\,^{(5)}G_{\alpha\gamma}(R)n^\alpha n^\gamma+\frac{3}{8}\,^{(5)}R\Big]+\frac{4}{3}q_{\beta\delta}\,^{(5)}R_{\sigma\nu}n^\sigma n^\nu \right. , \label{pro2}
\end{eqnarray} where $E_{\beta\delta}=C^\mu_{\nu\rho\sigma}n_\mu n^\rho q_\beta^\nu q_\delta^\sigma$. Now it is possible to write down the five-dimensional quantities by means of Eq. (\ref{1}). Hence Eq. (\ref{pro2}) reads
\begin{eqnarray}
^{(4)}G_{\beta\delta}(R)&=&\left.\frac{2}{3}\kappa_5^2\Big\{T_{\mu\nu}q_\beta^\mu q_\delta^\nu+q_{\beta\delta}(T_{\mu\nu}n^\mu n^\nu-\frac{1}{4}T)\Big\}-\frac{2}{3}{\cal K}_{\mu\nu}(q^\mu_\beta q^\nu_\delta+n^\mu n^\nu q_{\beta\delta})+\frac{5}{12}q_{\beta\delta}{\cal K}+ K K_{\beta\delta}-K_{\alpha\beta} K^\alpha_{\delta}\right.\nonumber\\&-&\left.\frac{1}{2}(K^2- K_{\alpha\gamma}K^{\alpha\gamma})q_{\beta\delta}-E_{\beta\delta}\right. .\label{pro3}
\end{eqnarray}

Imposing $\mathbb{Z}_2$ symmetry, a quite familiar orbifold characteristic of braneworld models \cite{RSI,Maa}, one has the behavior of the unitary orthogonal vector $n_\mu\mapsto -n_\mu$ when crossing the brane. As a matter of fact, since the extrinsic curvature is quadratic in the Gauss equation, the minus sign is not relevant. The complete expression for the extrinsic curvature is obtained by means of the appropriated junction conditions. Following a procedure akin to the one presented in (the Appendix of) Ref. \cite{Mc}, we shall think of the brane as a hypersurface orthogonally riddled by geodesics in such a way that the brane act as a truly interface between $r>0$ and $r<0$. In this vein, one is able to define the following brackets $[Q]=\lim_{r\rightarrow 0^+}Q-\lim_{r\rightarrow 0^+}Q$ for any tensorial quantity $Q$. Expressing, then, the relevant quantities by means of the Heaviside distribution, its derivatives and products must fulfill the rules of the distributional calculus, from which the Israel-Darmois junctions conditions arise. It is to be noticed, however, that ${\cal K}_{\mu\nu}$ does not have second derivatives in the metric and therefore both junction conditions are nothing else but the usual ones. Thus, attributing a energy-momentum tensor of the form 
\begin{equation}
T_{\mu\nu}=-\Lambda g_{\mu\nu}+S_{\mu\nu}\delta(r)+\tilde{T}_{\mu\nu}, \label{pro4}
\end{equation} the extrinsic curvature reads, as usual, $K_{\mu\nu}=-\frac{\kappa_5^2}{2}\Big(S_{\mu\nu}-\frac{1}{3}q_{\mu\nu}S\Big)$. In Eq. (\ref{pro4}) $\Lambda$ is the bulk cosmological constant, $S_{\mu\nu}$ the energy-momentum tensor on the brane, and $\tilde{T}_{\mu\nu}$ stands for any other eventual stress in the bulk. By its turn, $S_{\mu\nu}$ can also be decomposed into $-v q_{\beta\delta}+\tau_{\beta\delta}$ separating out the brane vacuum energy, $v$ (in the case of a homogeneous and isotropic brane), usually called the brane tension, from the stress-tensor on the brane, $\tau_{\beta\delta}$. Taking advantage of Eq. (\ref{pro4}) along with (\ref{pro3}) one arrive at the effective gravitational equation on the brane given by
\begin{eqnarray}
^{(4)}G_{\beta\delta}=-\Lambda_4 q_{\beta\delta}+\mathcal{F}_{\beta\delta}-\frac{2}{3}{\cal K}_{\mu\nu}q^\mu_\beta q^\nu_\delta+8\pi G_N\tau_{\beta\delta}-E_{\beta\delta}+\kappa_5^4\pi_{\beta\delta}, \label{pro5} 
\end{eqnarray} where  
\begin{equation}
\Lambda_4=\frac{\kappa_5^2}{2}\Big(\Lambda+\frac{\kappa_5^2}{6}v\Big)+\frac{{\cal K}}{4}-\frac{2}{3}{\cal K}_{\mu\nu}q^{\mu\nu},\label{pro6}
\end{equation}
\begin{equation}
\mathcal{F}_{\beta\delta}=\frac{2\kappa_5^2}{3}\Big\{\tilde{T}_{\mu\nu}q^\mu_\beta q^\nu_\delta+q_{\beta\delta}\Big(\tilde{T}_{\mu\nu}-\frac{\tilde{T}}{4}\Big)\Big\}, \label{pro7} 
\end{equation}
\begin{equation}
\pi_{\beta\delta}=-\frac{1}{4}\tau^\alpha_\beta\tau_{\alpha\delta}+\frac{1}{12}\tau\tau_{\beta\delta}+\frac{1}{8}q_{\beta\delta}\tau^{\alpha\gamma}\tau_{\alpha\gamma}-\frac{1}{24}q_{\beta\delta}\tau^2, \label{pro8}
\end{equation} and $G_N=\kappa_5^2 v /48\pi$.
There are many relevant points appearing in the Eqs. (\ref{pro5})-(\ref{pro8}). First let us evince the terms which are usual in the effective equations \cite{SSM}. The tensors $\mathcal{F}_{\beta\delta}$, $E_{\beta\delta}$, $\pi_{\beta\delta}$, and the effective Newton constant $G_N$ are the same of they counterpart when the projection starts from pure Einstein equation in five dimensions. In Eqs. (\ref{pro5}) and (\ref{pro6}) the novelty is, of course, encoded in ${\cal K}_{\mu\nu}$ and its trace. A special attention has to be paid to the Eq. (\ref{pro6}), which shows an interesting aspect arising in the implementation of braneworld models within such a non-conservative gravitational theory. This equation carries an effective cosmological constant which now becomes a function of the coordinates due to the terms inherited from the modified gravity. This feature is attractive from the cosmological point of view, as a time-dependent cosmological ``constant" makes possible a construction of a model of universe where the components of the dark sector are able to interact each other, exchanging energy and momentum \cite{int,int2,int3,int4,int5,int6,int7,int8,int9}. This class of cosmologies usually comes into play as an attempt of addressing the so-called ``coincidence problem" \cite{CP}. Obviously, in the well behaved limit of a vanishing ${\cal K}_{\mu\nu}$ the usual brane effective equations are recovered. An important characteristic appearing in the projected equations is shown when investigating the conservative law expressed in the Codazzi equation
\begin{equation}
D_\nu K^\nu_\mu-D_\mu K=\,^{(5)}R_{\rho\sigma}n^\sigma q^\rho_{\mu}, \label{pro9}
\end{equation} where $D_\mu$ is the covariant derivative with respect to $q_\mu\nu$. From (\ref{1}) it can be readily verified that 
\begin{equation}
^{(5)}R_{\rho\sigma}n^\sigma q^\rho_{\mu}=\kappa_5^2 T_{\rho\sigma}n^\sigma q^\rho_\mu-{\cal K}_{\rho\sigma}n^\sigma q^\rho_\mu, \label{pro10}
\end{equation} and hence Eq. (\ref{pro9}) gives 
\begin{equation}
D_\nu\tau^\nu_\mu=\Big(\frac{2}{\kappa_5^2}{\cal K}_{\rho\sigma}-\tilde{T}_{\rho\sigma}\Big)n^\sigma q^\rho_\mu. \label{pro11}
\end{equation} Notice that Eq. (\ref{pro11}) is to be analyzed in order to investigate the conservation of the brane stress-tensor. Usually, the existence of a non vanishing $\tilde{T}_{\mu\nu}$ is the responsible for the energy-momentum exchange between the brane and the bulk and, of course, for a null $\tilde{T}_{\mu\nu}$ the brane stress tensor is (covariantly) conserved. Here ${\cal K}_{\mu\nu}$ also shares this characteristic and even in the absence of $\tilde{T}_{\mu\nu}$ the non conservative gravity term act as the responsible for the brane-bulk energy-momentum exchange. It must be stressed, however, that the non conservative character of braneworld models with stresses in the bulk and of the gravity theory at hand may cancel each other, provided that
\begin{equation}
{\cal K}_{\mu\nu}=\frac{\kappa_5^2}{2}\tilde{T}_{\mu\nu}.\label{pro12}
\end{equation} These are a set of first order equations concerning bulk quantities (metric and energy-momentum content). This result comes exclusively from the non conservative gravity framework. 

We shall finalize this section pointing out that Eq. (\ref{pro12}) must be implemented for any (braneworld) model builder who want to ensure conservation of the brane stress tensor in the context studied here. It shall imprint a severe constraint on the model in question. In the next section we deserve more attention to this question, not by investigating a particular model, but instead appreciating the consequences of (\ref{pro12}) which are to be shared by any model constructed in such a scope.  

\section{Braneworld sum rules}

In trying to find out consistency conditions for braneworlds whose orbifold character is present, i. e., whose internal space is indeed compact, it was conceived an important formalism giving the necessary rules to be fulfilled by the plethora of models conceived since the publication of \cite{RSI}. This formalism was presented in Ref. \cite{KGL}, generalized in Ref. \cite{FR}, and studied under several different aspects \cite{JM}. We shall depict here the main relevant aspects for our purposes. When thinking of possible using the braneworld sum rules in the non-conservative gravity context a word of warning is in order. It seems possible, thought nontrivial, to find out the generalized partial traces coming from (\ref{1}) and thus to achieve the consistency conditions accordingly. Nevertheless, as we want to deal with non-conservative gravity theory in the bulk, we are going to use the standard protocol. 

For booking keep purposes we start with a $D-$dimensional bulk. Besides it is indeed more profitable to change the notation a bit making explicit the separation between bulk, brane, and internal space. The line element reads 
\begin{equation}
ds^2=g_{MN}dX^M dX^N=k_{mn}(r)dr^m dr^n+W^2(r)h_{\mu\nu}dx^\mu dx^\nu, \label{sm1}
\end{equation} where $M=\{m,\mu\}$ stands for the whole bulk index whose coordinates are denoted by $X^M$. The brane has $(p+1)$ dimension and is covered by coordinates $x^\mu$. Noticed that already we separated out the warp factor $W(r)$ contribution. Finally, the $(D-p-1)-$dimensional internal space is described by $k_{mn}$. Also, in order to make utterly clear the different geometrical quantities we denote by $\tilde{\tilde{A}}$ internal space quantities, while $\bar{\bar{A}}$ stands for a brane quantities. Thus it can be readily verified that 
\begin{eqnarray}
R_{\mu\nu}=\bar{\bar{R}}_{\mu\nu} -\frac{h_{\mu\nu}}{W^{p-1}}\nabla^2W^{p+1},\label{sm2}\\
R_{mn}=\tilde{\tilde{R}}_{mn}-\frac{p+1}{W}\nabla_m\nabla_n W, \label{sm3}
\end{eqnarray} where $\tilde{\tilde{R}}$, $\nabla_m$, and $\nabla^2$ are constructed out from $k_{mn}$. 

Now, with the aid of the partial traces $R^\mu_\mu=W^{-2}h^{\mu\nu}R_{\mu\nu}$ and $R^m_m=k^{mn}R_{mn}$ it is possible to write 
\begin{eqnarray}
\nabla \cdot (W^\alpha \nabla W)=\frac{W^{\alpha+1}}{p(p+1)}\Big[\alpha (\bar{\bar{R}}W^{-2}-R^\mu_\mu)+(p-\alpha)(\tilde{\tilde{R}}-R^m_m)\Big],\label{sm4}
\end{eqnarray} where $\alpha$ is a simple parameter, a freedom in the observance of the Leibniz rule $\nabla \cdot (W^\alpha \nabla W)=W^{\alpha+1}[\alpha W^{-2}\nabla W\cdot \nabla W+W^{-1}\nabla^2W]$. The values attributed to $\alpha$ at the end of the formalism shall give rise to a one-parameter family of consistency conditions. The key observation in applying the formalism in the context of non-conservative gravity is to derive the partial traces out from Eq. (\ref{1}). Hence we have 
\begin{eqnarray}
R^\mu_\mu=\frac{\kappa_5^2}{D-2}\Bigg((D-p-3)T^\mu_\mu-(p+1)T^m_m\Bigg)-{\cal K}^\mu_\mu,\label{sm5}\\
R^m_m=\frac{\kappa_5^2}{D-2}\Bigg((p-1)T^m_m-(D-p-1)T^\mu_\mu\Bigg)-{\cal K}^m_m,\label{sm6}
\end{eqnarray} where $K^\mu_\mu$ and ${\cal K}^m_m$ are defined as previously were their counterparts $R^\mu_\mu$ and $R^m_m$. They can be put in an explicit form as 
\begin{eqnarray}
{\cal K}^\mu_\mu=W^{-2}\lambda^\alpha(\partial^\mu h_{\mu\alpha}-h^{\mu\nu}\partial_{\alpha}h_{\mu\nu}),\label{sm7}\\
{\cal K}^m_m=\lambda^b\partial^m k_{bm}-\lambda^m k^{ab}\partial_m k_{ab}-2(p+1)\lambda^m\partial_m(\ln W).\label{sm8}
\end{eqnarray} The first two terms of (\ref{sm8}) shall eventually be discarded when making contact with Eq. (\ref{le}). We shall return to these equations latter. By now we remember that in a compact internal space the left-hand side of Eq. (\ref{sm4}) vanish upon integration. Therefore, taking back (\ref{sm5}) and (\ref{sm6}) into (\ref{sm4}), we have  
\begin{eqnarray}&&
\oint W^{\alpha+1}\left. \Bigg\{\alpha \bar{\bar{R}}W^{-2}-(p-\alpha)\tilde{\tilde{R}}+\alpha \tilde{K}^{\mu}_\mu+(p-\alpha)\tilde{{\cal K}}^m_m\right.\nonumber\\&+&\left. \frac{\kappa_5^2}{D-2}\Bigg(T^\mu_\mu[2\alpha+(D-p-1)(p-2\alpha)]+T^m_m p[2\alpha-p+1]\Bigg)\Bigg\}=0.\right.\label{sm9}
\end{eqnarray} 

The energy-momentum tensor (\ref{pro4}) may be suitable generalized to the sum rules formalism as 
\begin{equation}
T_{MN}=-\Lambda G_{MN}-\sum_{i}T^{(i)}P[G_{MN}]^{(i)}\Delta^{(D-p-1)}(r-r_i)+\tilde{T}_{MN},\label{sm10}
\end{equation} 
where, as usual in the sum rules formalism, $P[G_{MN}]^{(i)}$ denotes the pull-back of the bulk metric on the $i^{th}$-brane, $T^{(i)}$ is the tension of the $i^{th}$-brane and $\Delta$ is the generalization of the Dirac distribution localizing the brane in the internal space. These terms shall be simplified in the five-dimensional case. From (\ref{sm10}) $T^\mu_\mu$ and $T^m_m$ follows straightforwardly and, hence, Eq. (\ref{sm9}) may be recast in the form 
\begin{eqnarray}&&
\left. \oint W^{\alpha+1}\Bigg\{\alpha \bar{\bar{R}}W^{-2}+(p-\alpha)\tilde{\tilde{R}}-[\gamma+(D-p-1)\beta]\Lambda+\alpha {\cal K}^\mu_\mu+\frac{\gamma \kappa_5^2}{p+1}\tilde{T}^\mu_\mu+(p-\alpha){\cal K}^m_m\right.\nonumber\\&+&\left.\beta\kappa_5^2\tilde{T}^m_m-\kappa_5^2\gamma\sum_i T^{(i)}\Delta^{(D-p-1)}(r-r_i) \Bigg\}=0,\right. \label{sm11}
\end{eqnarray} where $\beta=\frac{p(2\alpha-p+1)}{D-2}$ and $\gamma=\frac{p+1}{D-2}[(p-2\alpha)(D-p-1)+2\alpha]$. Now it is possible to implement the particularizations we are interested, relating the formalism with the previous section. In this vein, we set $D=5$, $p=3$ leading immediately to $\tilde{\tilde{R}}=0$. Eq. (\ref{sm11}) then reads 
\begin{eqnarray}&&
\left. \oint W^{\alpha+1}\Big\{\alpha \bar{\bar{R}}W^{-2}-2\Lambda(\alpha+1)+\alpha{\cal K}^\mu_\mu+(3-\alpha){\cal K}^m_m+\kappa_5^2 \tilde{T}^\mu_\mu+2(\alpha-1)\kappa_5^2\tilde{T}^m_m\Big\}=4\kappa_5^2\sum_i T^{(i)}.\right. \label{sm12}
\end{eqnarray} Eq. (\ref{sm12}) provides a one parameter family of consistency conditions. Notice that in the appropriate limit $\lambda^M\rightarrow 0$ the usual sum rules are recovered \cite{FR}, as expected.

In trying to describe our universe in the four-dimensional brane, one is able to set $\bar{\bar{R}}=0$. From the plethora of possibilities arising from Eq. (\ref{sm12}) the condition coming from $\alpha=-1$ deserves to be highlighted. Usually, this choice reveals the necessity (or not) of a negative brane tension in the model. Using a standard bulk scalar field $\Phi(r)$ whose stress tensor reads 
\begin{eqnarray}
\tilde{T}_{\mu\nu}=-W^{-2}h_{\mu\nu}\Big(\frac{1}{2}\nabla\Phi\cdot\nabla\Phi +V(\Phi)\Big),\label{sm13}\\ 
\tilde{T}_{mn}=\nabla_m\Phi \nabla_n\Phi-k_{mn}\Big(\frac{1}{2}\nabla\Phi\cdot\nabla\Phi +V(\Phi)\Big),\label{sm14}
\end{eqnarray} it is possible to rewrite Eq. (\ref{sm12}) as 
\begin{eqnarray}
\oint \Big\{-{\cal K}^\mu_\mu+4{\cal K}^m_m-4\kappa_5^2(\nabla\Phi)^2 \Big\}=4\kappa_5^2\sum_i T^{(i)}, \label{sm15}
\end{eqnarray} from which we see the possibility of a smooth extension of the Randall-Sundrum model without the necessity of a negative brane tension. This is indeed an attractive aspect for braneworld modeling in this non-conservative framework.  

We are now in position to analyze from the sum rules perspective the peculiar output resulting from out previous section investigation, namely: the possibility of the (covariant) conservation of the brane stress-tensor provided that the constraint (\ref{pro12}) is verified. Notice that by taking advantage of Eq. (\ref{pro12}) we have 
\begin{equation}
4{\cal K}^m_m-{\cal K}^\mu_\mu=2\kappa_5^2(\nabla\Phi)^2 \label{sm16}
\end{equation} and, from (\ref{sm15}), it becomes clear the impossibility of the previously alluded smooth extension. Therefore, for braneworlds built under the auspices of the non-conservative gravity or one have conservation of the brane energy-momentum tensor or choose a non-negative brane tension context. 

We finalize pointing out that the sum rules might also be suitable to imprint some conditions on the $\lambda^A$ vector too. For instance, by working with a particular case in which $\lambda^A=(0,\lambda^r)$, then from Eqs. (\ref{sm7}) and (\ref{sm8}) we have ${\cal K}^\mu_\mu=0$ and ${\cal K}^m_m=-4\lambda^r\partial_r(\ln W)$. Hence (\ref{sm15}) leads to the following condition (again choosing $\alpha=-1$) 
\begin{equation}
\oint \Bigg\{4\lambda^r\frac{d(ln W)}{dr}+\kappa_5^2\Bigg(\frac{d\Phi}{dr}\Bigg)^2\Bigg\}=-\kappa_5^2\sum_i T^{(i)}, \label{sm17}
\end{equation} and a constant $\lambda^r$ would also make impossible to preclude from a negative brane tension (note that in this case the first integral in the left-hand side of (\ref{sm17}) indeed vanishes). It shows that the vector engendering the non-conservation may not be a completely free variable if one wish to study braneworld models without negative brane tension. We remark parenthetically that similar statements arise even given up of exclusively positive brane tension and use the constraint (\ref{pro12}) along with the bulk scalar field and the simple choice $\lambda^A=(0,\lambda^r)$. In fact, as ${\cal K}^\mu_\mu=0$ one is forced to conclude that $V<0$, a definitely odd scenario. 

\section{Conclusion}

The study we have performed exhausted the formal approach concerning braneworld scenarios based upon non-conservative gravity. It is important to emphasize that even in the most rudimentary approach, the result encoded in Eqs. (\ref{pro5})-(\ref{pro6}) is promising from the cosmological point of view. In fact, in a context in which the $\pi_{\mu\nu}$ can be disregarded (notice the $\kappa_5^4$ coefficient), the bulk has no additional stresses, and the geometrical set up carrying symmetries enough to set $E_{\mu\nu}=0$, the remaining effective field equation has some properties potentially interesting at cosmological level.    

We have analyzed a possible braneworld setup based upon a gravitational theory recently proposed where dissipative effects are introduced in the least action principle. We used this framework to generalize the consistency conditions to be obeyed by any viable braneworld model. We have shown that these non-conservative terms appearing in the new consistency relations open the possibility of relaxing the negative tension condition verified in the Randall-Sundrum context, so avoiding an undesirable property which plagues some braneworld models. Besides, we have seen through Eq. (\ref{pro6}) that this model of gravity provides a braneworld scenario with a running effective cosmological ``constant". As such this novel aspect is promising for cosmology as it can make feasible the emerging of interactions between dark energy and dark matter \cite{int,int2,int3,int4,int5,int6,int7,int8,int9}.

Our study also shows that the model investigated is allowed to have a standard conservation law for the energy-momentum tensor on the brane even with a non-zero stress in the bulk. On the other hand, we have seen that it is also possible to exist exchange of energy between the brane and the bulk, even if there is no stress in the bulk. The cosmological consequences of the possibilities arising in the present study shall be investigated in a future opportunity. 

\begin{acknowledgments}
JCF thanks to CNPq and FAPES for the financial support. JMHS has been partially supported by CNPq and also thanks to the Cosmology group of UFES for the kind hospitality. TRPC is grateful to CAPES for the full financial support. 
\end{acknowledgments}

\end{document}